\documentclass{emulateapj}

\usepackage{url}
\usepackage{amsmath}
\usepackage{amssymb}
\usepackage{mathrsfs}
\usepackage{natbib}
\usepackage[usenames,dvipsnames]{color}
\usepackage{ulem} 
\usepackage{amsfonts}
\usepackage{float}
\usepackage{graphicx}
\usepackage{verbatim} 
\usepackage{enumerate}

\def\draftmark {
}

\def\smbh#1{supermassive black hole{}#1
  (SMBH#1)\gdef\smbh{SMBH}}
\def\bh#1{black hole#1
  (BH#1)\gdef\bh{BH}}
\def\bbh#1{binary \bh{}#1
  (BBH#1)\gdef\bbh{BBH}}
\def\gw#1{gravitational wave#1
  (GW#1)\gdef\gw{GW}}
\def\imbh#1{intermediate mass black hole#1
  (IMBH#1)\gdef\imbh{IMBH}}
\def\ms#1{main sequence#1
  (MS#1)\gdef\ms{MS}}
\def\gc#1{galactic center#1
  (GC#1)\gdef\gc{GC}}
\def\CO#1{compact object#1
  (CO#1)\gdef\CO{CO}}
   \def\tde#1{tidal disruption event#1
  (TDE#1)\gdef\tde{TDE}}
   \def\HVS#1{hyper-velocity star#1
  (HVS#1)\gdef\HVS{HVS}}
\def\emri#1{extreme mass-ratio inspiral#1 (EMRI#1)\gdef\emri{EMRI}}
\def\bemri#1{binary extreme mass-ratio inspiral#1 (BEMRI#1)\gdef\bemri{BEMRI}}
\def\CBC#1{compact binary coalescence#1 (CBC#1)\gdef\CBC{CBC}}

\begin{document}

\def\M05{\citet{2005ApJ...631L.117M}}

\draftmark

\title{Busting Up Binaries: Encounters Between Compact Binaries and a Supermassive Black Hole}

\author{Eric Addison}
\affiliation{Department of Physics, Utah State University, Logan, UT 84322, USA.}

\author{Pablo Laguna}
\affiliation{Center for Relativistic Astrophysics and School of Physics, \\
School of Physics, Georgia Institute of Technology, Atlanta, GA 30332, USA.}

\author{Shane L. Larson}
\affiliation{Center for Interdisciplinary Exploration and Research in Astrophysics,\\ Northwestern University, Evanston, IL 60208, USA\\
and Department of Astronomy, Adler Planetarium, Chicago, IL 60605 USA.}

\begin{abstract}
    Given the stellar density near the galactic center, close
    encounters between compact object binaries and the supermassive
    black hole are a plausible occurrence.  We present results from a numerical study 
   of close to 13 million such encounters. Consistent with previous
    studies,  we corroborate that, for binary systems  tidally disrupted by the black hole,
   the component of the binary remaining bound to the hole has eccentricity $\sim 0.97$ 
and circularizes dramatically by the time
    it enters the classical LISA band. Our results also show that the population of surviving binaries merits attention.  These binary 
    systems experience perturbations to their internal orbital parameters with potentially interesting observational consequences. We investigated
    the regions of parameter space for survival and estimated the distribution of orbital
    parameters post-encounter.  We found that surviving binaries harden and their eccentricity increases, thus accelerating their merger due
    gravitational radiation emission and increasing the predicted
    merger rates by up to $1\%$.
\end{abstract}

\section{Introduction}\label{sec:Intro}

Observations of \gw{s} will allow us to probe dynamical
astrophysical systems in regimes where strong gravity plays a key role.
In galactic nuclei, \gw{}s from stellar mass interactions with a \smbh{} will be excellent probes of the properties of the
\smbh{} and the stellar population itself. It is thus important to investigate the
types of encounters that should be expected as well as the strength and frequency of \gw{s}
emitted by these events. 

In the present work, we will focus on encounters between stellar mass binaries and a
galactic \smbh{}. Previous studies have focused on the \HVS{} produced by the disruption of the binary
and stellar collisions
\citep{1988Natur.331..687H,2005MNRAS.363..223G,2010ApJ...713...90A}. 
\M05{} investigated the eccentricity of the bound component created by the binary disruption, namely the \emri{} left behind.  More recently, binary
interactions with \smbh{} have been explored in the context of the \gw{} emissions in
the LIGO band from binaries driven to merger by Kozai resonance
\citep{2012ApJ...757...27A, 2014ApJ...781...45A}.

The strongest \gw{} sources will involve compact, degenerate stellar
remnants that can survive close encounters with the central black
hole, namely neutron stars and stellar mass \bh{}s.  The fraction
of field stars in binaries varies by stellar type, but for O- and
B-type stars, which are the stellar types massive enough to form
neutron stars or \bh{}s, it is estimated that more than $75\%$ of
O-type and $70\%$ of B-type stars have some number of companions
\citep{2010ApJS..190....1R}.  Near the galactic center, the density of stars grows
very large compared to field conditions, with density estimates up to
$10^{8}M_{\odot} \text{ pc}^{-3}$ within the inner $0.1$pc,
\citep{2005PhR...419...65A}.  Given this information, it is reasonable
to expect the existence of  \CO{} binaries in the galactic center, and
in fact it has been observed by way of X-ray transients that \CO{}
binaries exist near the GC, and are more abundant within
the inner 1 pc \citep{2005ApJ...622L.113M}.  It has been estimated that as many as
$\sim 20,000$ stellar mass BH binaries have segregated within the
inner $\approx 1$ pc of the Milky Way galactic center \citep{2009MNRAS.395.2127O}.

With the existence of \CO{} binaries near the galactic center, it is reasonable to
expect that some number of them may interact directly with the \smbh{}.
Many known main sequence stars exist in bound orbits around the \smbh{}
\citep{2005ApJ...620..744G,2008ApJ...689.1044G,2009astro2010S..89G},
suggesting that the same could be true for \CO{}s and \CO{} binaries.  Such
interactions have implications for \gw{} campaigns of observations, for example ground based observatories (e.g. LIGO)
searching for \CBC{s}.  Proposed space based
interferometers (e.g. LISA) will be sensitive to \emri{}s which can be
created by the tidal disruption of \CO{} binaries near the \smbh{}.

In the present study, we corroborate and extend the results of
\M05{} for the formation of \emri{}s by tidal disruption to arbitrary binary
orientations. In addition, we focus our attention on those binary
systems that survive the encounter with the \smbh{}. In particular, we investigate
whether those binaries are able to survive and face subsequent encounters before merging from \gw{} emission or 
merge faster than those not experiencing encounters with \smbh{}, and discuss the
implication of the encounters  on  predicted
\CBC{} rates. We concentrate on initially circular, equal mass ($m_1 = m_2 \equiv m$), \CO{} binaries approaching the \smbh{} in a parabolic orbit.
We vary the orientation of the orbital angular momentum of the \CO{} binary relative to the angular momentum 
of the \CO{} binary orbiting the \smbh{}. We also vary the pericenter distance $r_p$ between the \CO{} binary and the \smbh{}.
Our study explores a greater volume of the
parameter space;
\citet{2012ApJ...757...27A} focused on binaries bound to the \smbh{} at a
fixed orientation, and \M05{} explored hyperbolic encounters with
coplanar binaries. Considering a larger region of parameter space comes at the expense of integration sophistication,
though the time and distance scales involved in our simulation suggest
that relativistic effects will not play a significant role for the
majority of the parameter range, and Newtonian gravitational forces
should suffice.

The paper is organized as follows:  
Section \ref{sec:analytic} outlines the outcomes of the three body
encounters based on energy arguments.  
In Section \ref{sec:Num}, we describe the setup of the encounters. 
Parameter probability distributions and an estimate of the tidal radius are given in 
section \ref{sec:tidal}. 
The analysis of the disrupted binaries is found in Section \ref{sec:disrupted}.
In Section \ref{sec:survive}, we discuss the results of those binaries that survived the encounter with the \smbh{.}
Section~\ref{sec:Peters} includes a discussion of the changes in the lifetime of the \CO{} binary due to \gw{} emission as a result of the encounter with the hole as well as consequences to \gw{} detection rates for \CO{} binaries.
The paper ends with conclusions in Section~\ref{sec:disc}.

\section{Energetics at a Glance}
\label{sec:analytic}

The total energy for a three-body system 
can be written as
\begin{equation}
   E = \dfrac{1}{2}\sum^3_{i=1}m_iv_i^2 - \sum_{i=1}^2\sum^3_{j=i+1}\frac{Gm_im_j}{|\bf{r}_i-\bf{r}_j|}\,\label{eq:Etot}
\end{equation}
In the barycentric frame, with $M_b = m_1+m_2$ the mass of the binary and $M_\bullet = m_3$ the mass of the \smbh{}, the total energy reads
\begin{equation}
   E = \dfrac{1}{2}\frac{M_b \,M_\bullet}{M_b+M_\bullet}{\bf V}^2 - \sum_{i=1}^2\frac{Gm_iM_\bullet}{|\bf{r}_i-\bf{r}_3|} + E_{b}\,.
\end{equation}
Above,
\begin{equation}
E_{b} = \frac{1}{2}\mu v^2 - \frac{G\mu M_b}{r} 
\end{equation}
 is the internal energy of the binary with ${\bf r} = {\bf r}_1 - {\bf r}_2$, ${\bf v} = {\bf v}_1 - {\bf v}_2$ and $\mu = m_1m_2/M_b$. Furthermore, 
\begin{equation}
{\bf V} = {\bf v}_3 - \frac{1}{M_b}\sum^2_{i=1} m_i{\bf v}_i
\end{equation}
is the relative velocity between the binary and the \smbh{}. Since $M_\bullet \gg M_b$, we can set  ${\bf r}_3 = 0$ and approximate ${\bf v}_3 \approx 0$. The total energy thus become 
\begin{equation}
   E = \dfrac{1}{2}M_b V_b^2 - \sum_{i=1}^2\frac{Gm_iM_\bullet}{r_i} + E_{b}\,.
\end{equation}
where $V_b$ is the velocity of the center-of-mass of the binary relativity to the \smbh{}. 
Since for the incoming binary $R \gg r$, with $R$ denoting the distance of the center-of-mass of the binary to the \smbh{},
we can rewrite the total energy as
\begin{equation}
   E \approx \dfrac{1}{2}M_b V_b^2 - \frac{GM_bM_\bullet}{R} + E_{b} = E_{cm}+E_b\,.\label{eq:CM}
\end{equation}
As mentioned before, we inject the binaries in parabolic orbits; therefore, the initial center of mass energy of the binary with the black hole is $E_{cm,0} =0$. Therefore, the total energy is given by the initial internal binding energy of the binary, $E = E_{b,0} <0$.

The effect of the encounter is to re-distribute the energy available, i.e. $E_{b,0}$, among the three bodies. If the binary survives, $E_b < 0$. Given that  
\begin{equation}
   E  = E_{cm} + E_{b} = E_{b,0}  <  0 \label{eqn:energyApprox}\,,
\end{equation}
the \CO{} binary after the encounter could be  bound  ($E_{cm} < 0$) or unbound ($E_{cm} > 0$) to the \smbh{}.

On the other hand, if the \CO{} binary does not survive ($E_{b} >0$), the separation $r$ of its components will grow; thus, one can neglect
in Eq.~(\ref{eq:Etot})
\begin{equation}
\frac{Gm_1m_2}{|\bf{r}_2-\bf{r}_1|} \approx 0\,,
\end{equation}
  and rewrite the total energy as
\begin{equation}
E = \sum^2_{i=1}\left(\dfrac{1}{2}m_iv_i^2 - \frac{Gm_iM_\bullet}{r_i}\right)= \sum_{i=1}^2 E_i\,\label{eq:Etot2}\,,
\end{equation}
where we have used again that ${\bf r}_3 = 0$ and ${\bf v}_3 \approx 0$. Therefore, for \CO{} binaries that are disrupted
\begin{equation}
   E  = E_1 + E_2 = E_{b,0}  <  0 \label{eqn:energyApprox}\,.
\end{equation}
The possible outcomes in this situation are 
both \CO{}s bound to the \smbh{} ($E_{1},E_{2}<0$), or one bound and the other unbound ($E_{i}<0<E_{j}$). 
Clearly, the case in which bound components are unbound ($0<E_{1},E_{2}$) is not possible.

We will refer to a \CO{} binary in a bound orbit around a \smbh{} 
as a \bemri{}, and will classify the aftermath of the binary from the encounter with the \smbh{} into one of the following four classes: 
\vspace{-5pt}
\begin{itemize}
\setlength{\itemsep}{-2pt}
\item DB: Disrupted binary. 
\item Long \bemri{}: Survived binary bound to \smbh{} with $\tau_{\rm gw}> P_{\bullet}$.
\item Short \bemri{}: Survived binary bound to \smbh{} with $\tau_{\rm gw}< P_{\bullet}$.
\item SU: Survived binary unbound to the hole.
\end{itemize}
In this classification, $\tau_{\rm gw}$ refers to the binary merger lifetime from \gw{} emission, the so-called Peters lifetime
\citep{1964PhRv..136.1224P}, and  $P_{\bullet}$ is the period of the bound binary around the \smbh{}. 
One of the main motivations of our study is to investigate how the probability of a binary
falling into one of these categories could alter the predicted \CBC{}.

The fate of a \CO{} binary encountering a \smbh{}  depends on its {\it penetration factor} $\beta$. The penetration factor is defined as the ratio
$\beta \equiv r_t/r_p$, where $r_t$ is the {\it tidal radius} and $r_p$ the distance of closest approach to the \smbh{}. 
In analogy with the common definition of tidal radius for the stellar disruption of stars by massive \bh{}s, we 
define the tidal radius $r_{t}$ as
\begin{equation}
   r_t \equiv \left (\dfrac{M_\bullet}{M_b}\right )^{1/3} a_0\label{eq:rtidal}\,,
\end{equation}
where $a_0$ is the initial semi-major axis of the \CO{} binary.
The radius $r_t$ is an approximation to the distance where within the tidal forces by the \smbh{} exceed the self-binding energy of the binary. The exact distance will depend on the orbital parameters and orientation of the binary~\citep{1991Icar...92..118H,
1992Icar...96...43H}. One of the objectives of our study is to provide a statistical estimate of the tidal radius from our set of encounters.

As mentioned before, the effect of the encounter is a redistribution of energy and angular momentum in the system.
As a consequence, orbital parameters in the \CO{} binary, such as eccentricity and semi-major axis are affected.
Regarding eccentricity, \citet{1996MNRAS.282.1064H} provided an analytic estimate of this perturbation for a circular binary system in parabolic orbit around a third body. For the case of $M_\bullet \gg M_b$, the perturbation reads
\begin{align}
   \delta e &= 3\sqrt{2\pi}\left(\frac{M_{b}}{M_{\bullet}}\right)^{1/4}\left
   ( \dfrac{2\, r_p}{a_0}\right )^{3/4}\notag \\
   &\times \exp \left [ -\dfrac{2}{3}\left
   (\dfrac{2M_{b}}{M_{\bullet}}\right )^{1/2}\left (\dfrac{r_p}{a_0}\right
   )^{3/2}\right ] {\cal F(\iota,\phi)}
   \label{eqn:H&R}
\end{align}
where
\begin{align}
 {\cal F(\iota,\phi)} &=  \cos^{2} \dfrac{\iota}{2}[\cos^{4} \dfrac{\iota}{2}+ \dfrac{4}{9}\sin^{4} \dfrac{\iota}{2}
  \notag\\
&  + \dfrac{4}{3}\cos^{2}
   \dfrac{\iota}{2}\sin^{2} \dfrac{\iota}{2}\cos \phi^{1/2}]\notag
\end{align}
with $\iota$
 the  inclination of the \CO{} binary relative to the orbit around the hole, and $\phi$ depending on
the initial phase of the binary and the longitude of the ascending
node.  
Inserting the definition of tidal radius,  Eq.~\ref{eqn:H&R} becomes
\begin{align}
   \delta e &= 6\,\sqrt{\pi}\,2^{1/4}\beta^{-3/4}\exp \left [ -\dfrac{2\,\sqrt{2}}{3}\beta^{-3/2}\right ] {\cal F(\iota,\phi)}
   \label{eqn:H&Rbeta}
\end{align}
This expression will be compared against our simulation results in
section \ref{sec:ecc}.  Since $\delta e$  vanishes as inclination
$\iota \rightarrow \pi$ and $\beta  \rightarrow 0$, we expect to see stronger agreement
between simulation results and Eq.~\ref{eqn:H&Rbeta} for the low
inclination and large pericenter encounters.

\section{Binary Encounter Setup}
\label{sec:Num}

The \CO{} binary is injected in a parabolic orbit around
the \smbh{}. The integration runs
until either the binary is tidally disrupted by the \smbh{} and an amount
of time equal to the initial time passes, or the center of mass of the
binary has reached a true anomaly of $\Theta = -\Theta_{0}$, where
$\Theta_{0}<0$ is the initial true anomaly of the orbit of the binary's center of mass about the \smbh{}.

The following parameters in the three-body system are kept fixed: \smbh{} mass ($M_\bullet = 10^6\,M_\odot$), \CO{} binary masses ($m_1 = m_2 = m = 10\,M_\odot$), initial binary eccentricity ($e_0 = 0$), and initial binary semi-major axis ($a_0 = 10\,R_\odot = 0.047$ AU). With these parameters, the \CO{} binary has a period 
\begin{eqnarray}
P_{b,0} &=& \frac{2\,\pi\,a_0^{3/2}}{\sqrt{G\,M_b}}\nonumber\\
              &=& 7.16\times 10^4\,{\rm s}\left(\frac{a_0}{10\,R_\odot}\right)^{3/2} \left( \frac{M_b}{20\,M_\odot}  \right)^{-1/2}\,,
\end{eqnarray} 
and tidal radius of the three-body system is
\begin{eqnarray}
        r_t &=& 173\, r_g\,M_6^{-2/3}\left(\frac{M_b}{20\,M_\odot}\right)^{-1/3}\left( \frac{a_0}{10\,R_\odot}\right)\nonumber\\
            &=& 1.73\,{\rm AU}\,M_6^{1/3}\left(\frac{M_b}{20\,M_\odot}\right)^{-1/3}\left( \frac{a_0}{0.047\,{\rm AU}}\right)\,.  \label{eqn:rt}
\end{eqnarray}
where $M_6 \equiv M_\bullet/10^6\,M_\odot$ and $r_g = G M_\bullet/c^2$ is the gravitational radius. 

The parameters we vary are the \CO{} binary orbital
inclination $\iota$, the initial longitude of the ascending node
$\Omega_0$, the binary
initial phase $\theta_0$, and the pericentric distance $r_p$ via the penetration factor $\beta$. 
Thus, our simulations of \CO{} binary encounters with a \smbh{} span a four-dimensional parameter space
$\lbrace \beta, \iota, \Omega_0, \theta_0 \rbrace$. We sample this parameter space with random values taking uniform 
distributions in $\beta^{-1} \in [0.35, 5]$, $\cos \iota \in[-1, 1]$, and $\Omega_0 \in [0, 2\pi]$. For the phase $\theta_0$, we take $200$ evenly spaced values
between $0$ and $2\pi$. The distribution in $\beta^{-1}$ implies that the values of the pericentric distance are uniformly distributed in
$r_t \in [0.865,8.65]$ AU. Furthermore, since the duration of the encounter is of order
\begin{eqnarray}
T_p &=& \frac{2\,\pi\,r_p^{3/2}}{\sqrt{G\,M_\bullet}} = \beta^{-3/2}\,P_{b,0}\,,
\end{eqnarray}
the range of parameters we consider for $\beta^{-1}$ imply that $T_p \in [0.1,3]\,P_{b,0}$. Moreover, in the nomenclature of Heggie, our binary system is ``soft'' since
\begin{equation}
\frac{1}{2}M_bV_p^2 \gg -E_{b,0} 
\end{equation}
which can be seen from
\begin{eqnarray}
\frac{1}{2}M_bV_p^2 &=& \frac{G\,M_b\,M_\bullet}{r_p}= \frac{G\,M_b^2}{a_0} \left(\frac{M_\bullet}{M_b}\right)^{2/3}\beta\nonumber \\
&\gg& \frac{G\,M_b^2}{4\,a_0} = -E_{b,0} \,.
\end{eqnarray}
Also, since $r_t \sim 173\,r_g$, even for the deepest penetration encounters with $\beta = 2$, the pericentric distance will be $r_p = 87\,r_g$. Therefore, it is perfectly safe to ignore general relativistic effects in all our encounters.
Similarly, given that at pericentric passage
\begin{equation}
\frac{v^2}{c^2} \sim \frac{G\,M_\bullet}{c^2\,r_p} \sim \frac{1}{87}\,,
\end{equation}
post-Newtonian corrections to the orbital motion of the binary are at the level of a percent and thus will be ignored. 
However, in analyzing the the aftermath of an encounter, we will take into account the Peters lifetime $\tau_{gw}$ of the binary, if survived, or if the \emri{} if the binary is disrupted.
For the \CO{} binary systems that we consider, the Peters lifetime due to the emission of \gw{s} in the absence of the \smbh{} is \citep{1964PhRv..136.1224P}
\begin{equation}
\label{eq:pedro}
\tau_{{\rm gw},0} = 0.95\times 10^{8} \,{\rm yr} \left(\frac{a_0}{10\,R_\odot}\right)^4\left(\frac{M_b}{20\,M_\odot}\right)^{-3}\,.
\end{equation}

The present study consists of $N_{c} \approx 13$ million individual simulations, resulting in: 2.1 million simulations of DB type encounters,
1.7 million yielding long \bemri{}, 2 million producing short \bemri{s} and 7.1 million of the SU type. 
We integrate the Newtonian equations of motion for the three-body system 
using Burlish-Stoer extrapolation with a leap-frog
integrator as described in \cite{1999MNRAS.310..745M}. We increase the
accuracy of the integration by the use of the CHAIN concept ala
\cite{1993CeMDA..57..439M}.  The coordinate system used for the
numerical integration are barycentric coordinates with the three body
center of mass at the origin.  The orbital angular momentum of the \CO{} binary--\smbh{} system is aligned with the $z$-axis. That is, the plane of the \CO{} binary center of mass orbiting the \smbh{} is the $xy$-plane. We inject the \CO{} binary in the first quadrant ($x,y>0$)  at a distance $200\,r_{p}$ and orient the orbit of the \CO{} binary about \smbh{} 
such that pericentric distance $r_p$ occurs along the $x$-axis, specifically at $y=0$ and $x<0$. 
Conservation of energy and angular momentum is checked at every time
step, and simulations are halted and rejected if either quantity
deviates from the initial value by one part in $10^{6}$.

\section{Parameter probability distriputions and tidal radius}
\label{sec:tidal}

\begin{figure}
\centering
\includegraphics[width = \columnwidth]{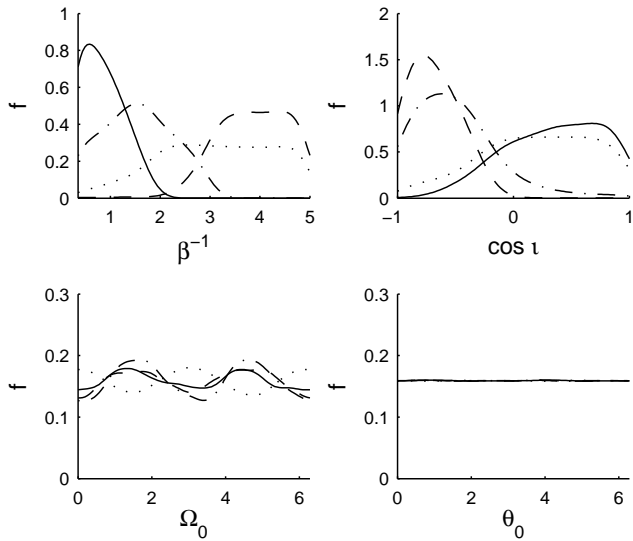}
\caption{KDE probability distribution function $f$ as computed from Eq.~\ref{eq:KDE} for each of the $\lbrace \beta, \iota, \Omega_0, \theta_0 \rbrace$ parameters. The pdf $f$ represents the probability that a binary with a particular end-state began with a particular value of a certain parameter. The lines-class correspondence is: solid denotes class DB, dot-dash class long \bemri{}, dash class short \bemri{}, and fine dash class SU.}
\label{fig:densEst}
\end{figure}

As already mentioned, we have classified the encounters into four types: 
DBs, long and short \bemri{s} and SUs. Next, for each of these outcome types, we 
estimate the probability distribution as a function of the varying parameters $\lbrace \beta, \iota, \Omega_0, \theta_0 \rbrace$. For example, $f(\beta^{-1} | DB)$ is the probability distribution function for the parameter $\beta^{-1}$ for disrupted binaries, i.e. the probability that a disrupted binary had a particular value of $\beta^{-1}$ (solid curve in upper left panel of Figure \ref{fig:densEst}).
We use for this purpose the technique of kernel density estimation (KDE).  KDE is a non-parametric method
for estimating probability densities in which a kernel function $K$
is convolved with a collection of Dirac delta functions. KDE
asymptotically converges to the true distribution faster than
histograms \citep{1979_KDE}. The KDE probability distribution $f$ of a parameter $x$ is computed from
\begin{equation}
\label{eq:KDE}
   f(x) = \dfrac{1}{N}\sum_{n=0}^{N} K(x) * \delta(x-x_n) =
   \dfrac{1}{N}\sum_{n=0}^{N} K(x-x_n)
\end{equation} 
where $N$ is the number of data points in the sample. We use a Gaussian
kernel $K(x) = (2\pi h^2)^{-1/2} \exp \left [ x^2/(2h^2)\right ]$ with
a variance of $h^2 = \text{[parameter range]}/100$, chosen to produce
distributions that retain structure while not being over-smoothed.
Figure~\ref{fig:densEst} shows the resulting normalized KDE probability distributions $f$ for each of the $\lbrace \beta, \iota, \Omega_0, \theta_0 \rbrace$ parameters obtained from Eq.~\ref{eq:KDE} .

We can draw from these probability distributions several conclusions about
the nature of the end-state of the binary relative to the input parameters,
as well as the predictive power of the individual parameters.  It is
clear from the panel of the $\beta^{-1}$ probability distribution (top left panel in Figure~\ref{fig:densEst})
that the encounters yielding binary disruption (i.e. DB-type, black, dot-dash line) occur primarily for small
values of $\beta^{-1}$ or large penetration factors $\beta$.  This should be obvious: closer passes translates into stronger tidal forces and higher likelihood of disruption.
Additionally, regarding the parameter $\iota$, prograde binaries (defined here as those with
inclination $\iota < \pi/2$) are more probable to disrupt than
retrograde binaries (those with $\iota > \pi/2$).  This is so because, in general,
retrograde binaries are more likely to survive and become bound to the
\smbh{}, implying that while the \bh{} orbit loses energy, the
surviving retrograde binaries must gain an equal amount energy, though
not generally enough to disrupt.   

\begin{figure}
\centering
\includegraphics[width = \columnwidth]{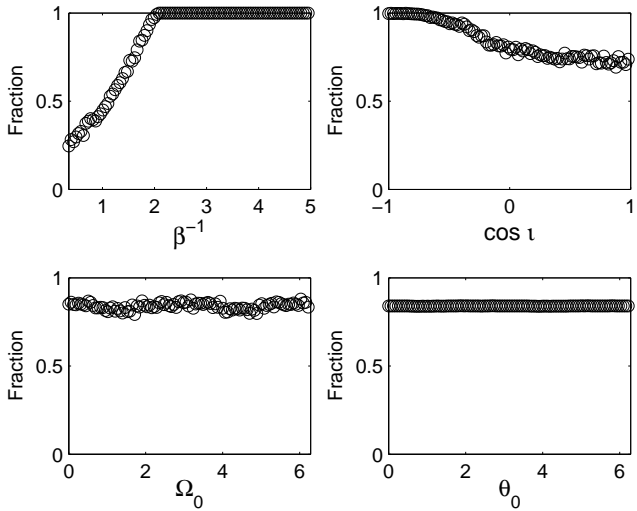}
\caption{Fraction of surviving binaries as a function of each in parameter for each parameter $\lbrace \beta, \iota, \Omega_0, \theta_0 \rbrace$.}
\label{fig:resultsDisPlot}
\end{figure}

Notice that the probability distribution for the initial orbital phase $\theta_0$ is flat. That is, the value of $\theta_0$ does not play a role on the outcome of the encounter. The distributions in Figure \ref{fig:resultsDisPlot} are over all the simulations in the parameter space. The flat distribution in $\theta_0$ illustrates that there is no strong correlation between $\theta_0$ and disruption across the entire parameter space of interest; the combination of other parameters has far greater influence. Clearly if all parameters except $\theta_0$ were fixed, some binaries would be disrupted, but there is no consistent trend.
Also interesting is that the probability of longitude
of the ascending node $\Omega_0$ is relatively flat. Therefore, we will focus our attention 
on the parameters $\beta$ and $\iota$ since they have the largest
effect on the binary end state.

The fact that $\beta$ and $\iota$ are the most relevant parameters can also be seen in Figure \ref{fig:resultsDisPlot} 
where we show the
fraction of surviving binaries for each parameter.  Consistent with Figure~\ref{fig:densEst}, the top panels in Figure \ref{fig:resultsDisPlot} show that 
the survival probability depends more strongly on $\beta^{-1}$ and $\iota$. The
top right panel shows again that  prograde ($\iota < \pi/2$) binaries in general more likely to be disrupted
than retrograde ($\iota > \pi/2$) binaries.  Similarly, it is clear from the top left panel that  the
probability of disruption goes to zero for $\beta^{-1} \gtrsim 2.1$.
Also consistent with Figure~\ref{fig:densEst}, the bottom panels in Figure \ref{fig:resultsDisPlot} show that $\Omega_0$ and $\theta_0$ have
a very weak influence on the survival ratio. Therefore, we can assume with confidence that for practical purposes the parameter space is
two-dimensional, namely $\lbrace \beta, \iota\rbrace$.

\begin{figure}
\centering
\includegraphics[width = \columnwidth]{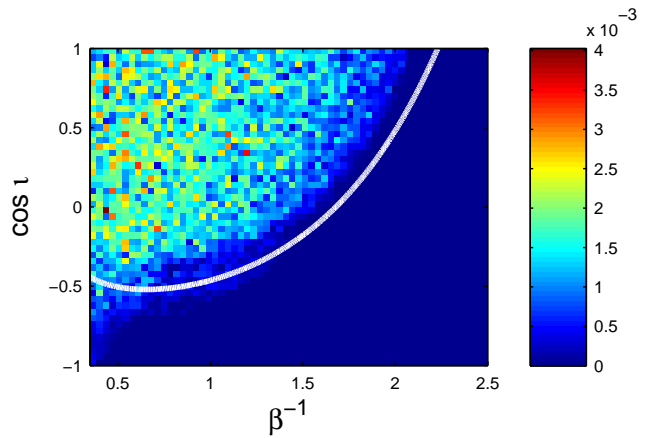}
\caption{2D histogram of disruptions vs. $\beta^{-1}$ and $\cos{\iota}$, shown as fraction of total disruptions. The white line denotes Eq.~\ref{eqn:H&Rbeta2}.}
\label{fig:resultsDisHistAnal}
\end{figure}

The dependence of the binary survival on $\iota$ implies that the tidal radius does not only depend on $\beta$ but also
on the inclination of the binary. This is not surprising because the forces from the \smbh{} responsible for disrupting the binary are those projected
along the semi-major axis of the binary. 
Figure \ref{fig:resultsDisHistAnal} depicts a two-dimensional histogram of tidal disruptions as a 
function of $\cos{\iota}$ and $\beta^{-1}$.  The solid line is the condition obtained by setting in Eq.~\ref{eqn:H&R} the eccentricity perturbation to $\delta e = 1$ with $\phi = 0$ to maximize the effect.  Namely,
\begin{align}
 {\cal F}^{-1}(\iota,0)  &= 6\,\sqrt{\pi}\,2^{1/4}\beta^{-3/4}\exp \left [ -\dfrac{2\,\sqrt{2}}{3}\beta^{-3/2}\right ] 
   \label{eqn:H&Rbeta2}
\end{align}
Notice that this
analytic approximation bounds the region containing disruptions well
for $\beta^{-1} \gtrsim 1$, but does not do so well for $\beta^{-1} <
1$, where the deepest penetration encounters are located.

\begin{figure}
\centering
\includegraphics[width = \columnwidth]{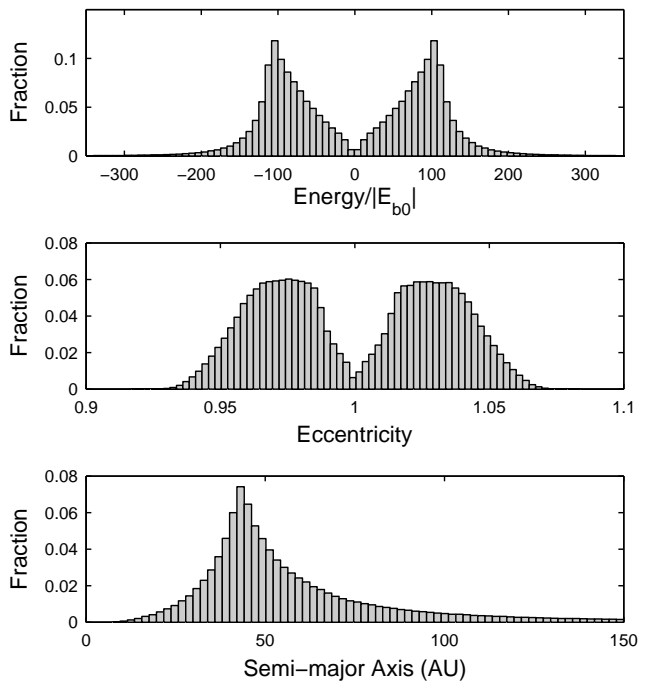}
\caption{Top panel: Histogram of the binding energy $E_i$ from Eq.~\ref{eq:Ei} normalized to the initial binding energy of the binary $E_{b,0}$. Middle panel:  Histogram of the corresponding eccentricity. Bottom Panel: Histogram of resulting EMRI semi-major axis. Histograms shown as fraction of total disruptions.}
\label{fig:resultsDisHist}
\end{figure}

\section{Disrupted Binaries}
\label{sec:disrupted}

As stated in the introduction, \CO{} binaries disrupted by the \smbh{} are of interest as both sources of \emri{}s and \HVS{}s. 
In this section, we investigate the channels for 
\emri{} formation and use previous \HVS{} results as a
check for our simulation accuracy.

We recall that after a binary is disrupted, the total energy of the system is approximately given by $E = E_1+E_2 = E_{b,0}$, where $E_{b,0}$ is the initial binding energy of the \CO{}  binary and
\begin{equation}
E_i = \dfrac{1}{2}m_iv_i^2 - \frac{Gm_iM_\bullet}{r_i}\label{eq:Ei}\,,
\end{equation}
with $i = 1,2$.

Figure~\ref{fig:resultsDisHist} shows the combined histograms of the normalized energies $E_{1}/|E_{b,0}|$ and $E_2/|E_{b,0}|$ (top panel) and the corresponding eccentricities (bottom panel). 
The symmetry of the histograms gives the impression when a binary is disrupted the outcome is invariable one component 
bound to the \smbh{}, i.e. an \emri{}, and the other component unbound to the hole, namely a \HVS{}. This is indeed the case for most of the disruptions ($\sim99.9967\%$). However, since $E_1+E_2 = E_{b,0}<0$, there are cases, although rare ($\sim0.0033\%$), for which both $E_1<0$ and $E_2<0$, and the outcome is two \emri{}s. Moreover, notice from the bottom panel in Figure~\ref{fig:resultsDisHist} that the average
eccentricity  for the bound (unbound) component of the disrupted binary is $e_- \approx 0.97$ ($e_+ \approx 1.3$). This can be understood from the definition of eccentricity
\begin{equation}
e_\pm^2 = 1 + \frac{2\,E_\pm\,L_\pm^2}{G^2\,m^3_\pm\,M^2_\bullet}\,,
\end{equation}
where we approximate
\begin{equation}
E_\pm \approx \pm G\,M_\bullet\,m_\pm\,a_0\,r^{-2}_p\label{eq:Ei}
\end{equation}
and
\begin{equation}
L^2_\pm \approx L^2_{\rm cm} = 2\,G\,M_\bullet\,m_\pm^2\,r_p\,.
\end{equation}
Thus, 
\begin{eqnarray}
e_\pm^2 -1 &=& \pm 4\frac{a_0}{r_p} = \pm 4\,\beta\frac{a_0}{r_t}= \pm 4\,\beta\left(\frac{M_b}{M_\bullet}\right)^{1/3}\nonumber\\
e_\pm &\approx& 1\pm 0.05\,\beta\left(\frac{M_b}{20\,M_\odot}\right)^{1/3}M_6^{-1/3}\,.
\end{eqnarray}
Furthermore, from Eq.~\ref{eq:Ei} and $E_{b,0} = |G\,M_b^2/(4\,a_0)|$, we have that 
\begin{equation}
\frac{E_\pm}{|E_{b,0}|} \approx \pm 74\,\beta^2\,M_6^{1/3}\,\left(\frac{M_b}{20\,M_\odot}\right)^{-1/3}\,,\label{eq:Ei2}
\end{equation}
consistent with the histograms in the top panel of Figure~\ref{fig:resultsDisHist}.

\subsection{Ejected Hyper-velocity Stars}
Hypervelocity stars are stars with velocities of the order of
hundreds or thousands of km s$^{-1}$, which may exceed the escape
velocity of our galaxy.  These stars were predicted by
\cite{1988Natur.331..687H} as the result of binary disruption in the
galactic center, and discovered observationally nearly two decades later
\citep{2005ApJ...622L..33B,2005ApJ...634L.181E}.

The production of \HVS{} from the tidal disruption of binary systems by a
\smbh{} has been discussed by \citet{1988Natur.331..687H,2003ApJ...599.1129Y,2010ApJ...713...90A,2006ApJ...653.1194B}.  
Our study is consistent with those results.
From Eq.~\ref{eq:Ei} and $E_{i} = m_i\,v_{\infty}^2/2$,  one has that
\begin{eqnarray}
v_\infty^2 &\approx& 2\,G\,M_\bullet\,a_0\,r^{-2}_p \nonumber\\
&\approx&  2\,G\,\beta^2\,M_\bullet\,a_0\,r^{-2}_t \nonumber\\
&\approx&  2\,G\,\beta^2\,M^{1/3}_\bullet\,a_0^{-1}\,M_b^{2/3}\,,
\end{eqnarray}
which yields the following asymptotic velocity approximation of a \HVS{}
 \begin{eqnarray}
v_{\infty} &\approx& 5,312\, \text{ km/s}\,\cdot\beta\,\cdot M_6^{1/6}\notag\\
&&\quad\times\left(\frac{a_0}{0.047\,\text{AU}}\right)^{-1/2}\left(\frac{M_b}{20\,M_\odot}\right)^{1/3}\,.
\end{eqnarray}
Rescaled to our case, the numerical study by \citet{2006ApJ...653.1194B} found that 
\begin{eqnarray}
\label{eq:hvsevj}
  v_{\infty} &\approx& 4,468  \text{ km/s}\,\cdot g(D) \,\cdot M_6^{1/6}\notag\\
&&\quad \times \left ( \dfrac{a_{0}}{0.047\text{AU}}\right )^{-1/2}\left
   (\dfrac{M_b}{20\,M_{\odot}}\right )^{1/3} 
\label{eqn:hvsevj}
\end{eqnarray}
where
\begin{eqnarray}
   g(D) &=& 0.774 + 0.0204\,D -6.23\times10^{-4}\,D^2 \notag\\
&+& 7.62\times   10^{-6}\,D^3 -4.24\times 10^{-8}\,D^4 \notag\\
&+&  8.62\times 10^{-11}\,D^5.
\end{eqnarray}
with $D = 79.37\,\beta^{-1}$.

\begin{figure}
\centering
\includegraphics[width = \columnwidth]{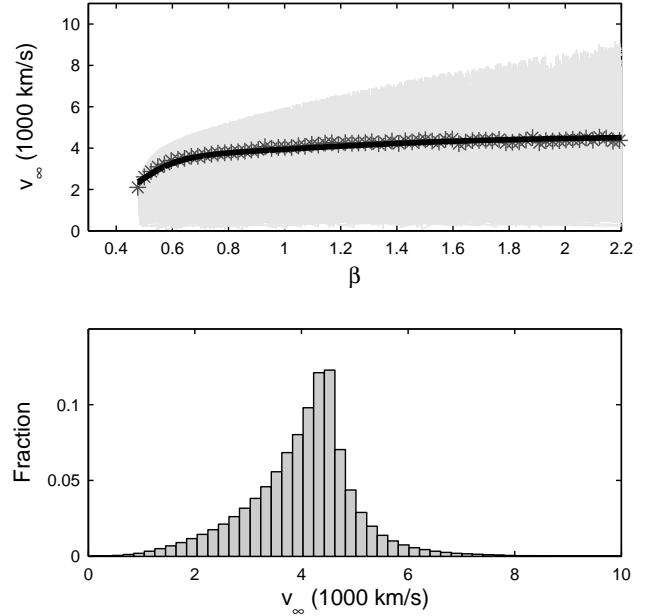}
\caption{Top panel: Grey dots denoted the ejection velocity $v_\infty$ as a function of $\beta$ for all the unbound stars from disrupted binaries. Stars show the averages over 100 evenly spaced intervals in $\beta$. Solid line shows the model prediction from Eq. \ref{eq:hvsevj}. Bottom panel: distribution of $v_{\infty}$ for all ejected components.}
\label{fig:resultsHVS}
\end{figure}

In the top panel in Figure~\ref{fig:resultsHVS} we show with grey dots  the ejection velocity $v_\infty$ as a function of $\beta^{-1}$ for all the unbound stars from disrupted binaries. In the same panel, stars show the averages over 100 $\beta$ bins. The solid line shows the model prediction from Eq. \ref{eq:hvsevj}. The agreement of our average velocities with those from \citet{2006ApJ...653.1194B} is evident. 
However, it should be noted
that the range of possible velocities can vary significantly from this
average, with the largest values roughly double the analytic
prediction and nearly $59\%$ of ejections in our simulations exceeding
the predicted value.

\subsection{Extreme-Mass-Ratio Inspirals}

\begin{figure*}
\centering
\includegraphics[width = \textwidth]{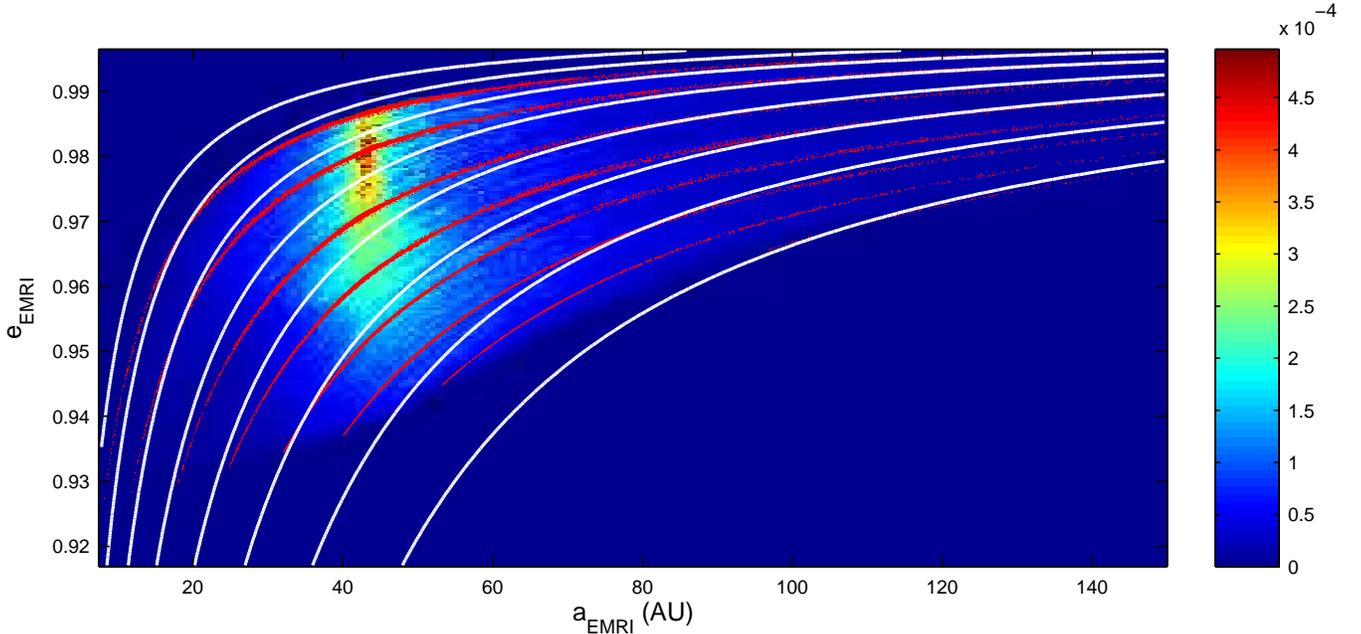}
\caption{2D histogram of potential  \emri{} candidates as a function of eccentricity $e_i$ and semi-major axis $a_i$, shown as fraction of total disruptions. White lines are lines of constant Peters lifetime with values $\tau_{\rm gw} = 10^{\alpha}$ years, where from left to right $\alpha =$ 5.0,  5.5, 6.0, 6.5, 7.0, 7.5, 8.0 and 8.5, respectively. Red lines show lines of constant $\beta^{-1}$ with $\beta^{-1} =$ 0.35, 0.5, 0.75, 1.0, 1.25 1.5 1.75, and 2.0 from left to right.}
\label{fig:resultsEMRIheat}
\end{figure*}

The traditional formation channel of an \emri{} is when two-body relaxation not only bounds a star to a \smbh{} but also brings it into the \gw{-}emission inspiriling regime \citep{0264-9381-24-17-R01}. The pericentric distance to capture a star bound to a \smbh{} with semi-major axis $a_c$ and form an \emri{} is 
\begin{equation}
r_c \approx 3\,r_g M_6^{-4/7}\left(\frac{m}{10\,M_\odot}\right)^{2/7}\left(\frac{a_c}{0.05\,{\rm pc}}\right)^{2/7}\,.
\end{equation} 
Since the star needs to penetrate so deeply into the
the neighborhood of the \smbh{,} despite circularization from \gw{} emission,
the \emri{} arrives to merger with significant eccentricity, 
$e\sim 0.5-0.9$.

The disruption of a binary by a \smbh{} provides an alternative channel since the disruption will always
leave at least one component bound to the \smbh{.} The capture distance in this situation is much larger. It is not $r_c$ but
the tidal radius $r_t \sim 173\,r_g$ (see Eq. \ref{eqn:rt}). Being bound to the \smbh{} does not necessarily translates into becoming an \emri{.}
The condition for a star to qualify as an \emri{} is that the timescale $\tau_{\rm gw}$ for its orbital decay by GW emission is sufficiently shorter than the relaxation time 
$t_{\rm rlx}$ of the stars in the galactic center  \citep{0264-9381-24-17-R01}:
\begin{eqnarray}
t_{\rm rlx} &\approx& 1.8\times 10^8\,{\rm yr} \left(\frac{\sigma}{100\,{\rm km}\,{\rm s}^{-1}}\right)^3\notag\\
&\times&\left( \frac{10\,M_\odot}{m}\right) \left(\frac{10^6M_\odot{\rm pc}^{-3}}{\bar m \, n}\right)
\end{eqnarray}
where $\sigma$ is the local velocity dispersion, $n$ is the local number density of stars, and $\bar m $ is the average stellar mass.

Figure \ref{fig:resultsEMRIheat} displays a 2D histogram  of $(a,e$) for \emri{} candidates from the disruption of binaries. White lines are lines of constant Peters lifetime with values $\tau_{\rm gw} = 10^{\alpha}$ years, where from left to right $\alpha =$ 5.0,  5.5, 6.0, 6.5, 7.0, 7.5, 8.0 and 8.5, respectively. Red lines show lines of constant $\beta^{-1}$ with $\beta^{-1} =$ 0.35, 0.5, 0.75, 1.0, 1.25 1.5 1.75, and 2.0 from left to right. 
It is clear from this histogram that most of the candidates qualify as \emri{s} since $\tau_{\rm gw} \ll t_{\rm rlx} \approx 10^8$ yr.
Also interesting is that the encounters with deepest penetration factors, i.e.  small $\beta^{-1}$, yield \emri{s} with larger eccentricity.

\begin{figure}[t]
\centering
\includegraphics[width = \columnwidth]{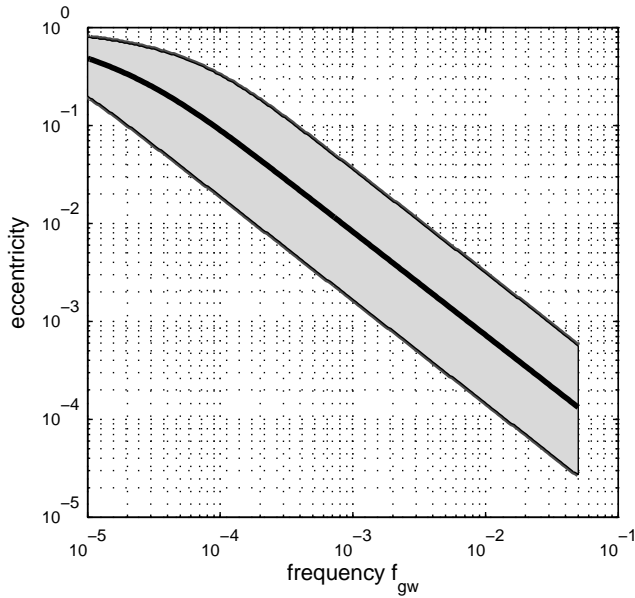}
\caption{\emri{} eccentricity at fundamental frequency $f_{\rm gw}$. Eccentricity for individual \emri{}s are shown in gray, average shown in black.}
\label{fig:resultsLISAf}
\end{figure}

Our results are consistent with those from \M05{} in the following respect. Because their ``capture'' radius is much larger than in the
traditional scenario, i.e. $r_t \sim 170\,r_g$ instead of $r_c \sim {\rm few}\, r_g$, the \emri{} from binary disruptions will have circularized dramatically by the time they merger or they enter the sensitive band of a LISA-like detector.  One can see this using~\citet{1964PhRv..136.1224P}
\begin{equation}
\label{eq:peters_a}
   a(e) = \dfrac{c_0\,e^{12/19}}{(1-e^{2})}\left ( 1 +
   \dfrac{121}{304}e^{2} \right )^{870/2299}
\end{equation}
with $c_0$ determined by the initial condition $a=
a_0$ when $e=e_0$. From Figure~\ref{fig:resultsEMRIheat}, we have that  at ``birth'' the \emri{} will have a typical eccentricities of $e_0 \approx 0.97$ and semi-major axis $a_0 \approx 45$ AU; thus, $c_0 \approx 2.3 \text{ AU}$ in Eq.~\ref{eq:peters_a}. For a space-based interferometer like LISA, the low end of the sensitivity window is $f_{\rm gw} \sim 10^{-4}\,\text{ Hz}$. If we approximate $f_{\rm gw} \approx 2\,f$ with $f^{-1} = P = 2\,\pi\,a^{3/2}/\sqrt{G\,M_\bullet}$ the Keplerian orbital period, we have that when the \emri{} enters the sensitivity band of a LISA-like detector it has a semi-major axis of
\begin{equation}
a \lesssim 74\,r_g\left(\frac{f_{\rm gw}}{10^{-4}\,\text{Hz}}\right)^{-2/3}\,M_6^{-2/3}\,.
\end{equation}
From Eq.~\ref{eq:peters_a}, one has that the corresponding eccentricity is $e \lesssim 0.15$. By the time the \emri{} merges, i.e. $a \sim r_g$, it will have eccentricity $e\sim 10^{-4}$ and emit \gw{s} at dominant frequencies of  $f_{\rm gw} \approx 60 \text{ mHz}$. 
This is clear from Figure~\ref{fig:resultsLISAf} where we plot the 
eccentricity of the \emri{s} as a function of the \gw{} frequency $f_{\rm gw}$. We approximate $f_{\rm gw} \approx 2\,f$ with $f^{-1} = P = 2\,\pi\,a^{3/2}/\sqrt{G\,M_\bullet}$ the Keplerian orbital period and $a$ from \citep{1964PhRv..136.1224P}.
Figure \ref{fig:resultsLISAf}
shows that in the sensitive LISA band
($0.1\text{ mHz} \lesssim f_{\gw{}} \lesssim 100\text{ mHz}$),  \emri{}s can
have average eccentricity $\bar e \le 0.1$.  At
the most sensitive frequencies, however, at $f_{\gw{}} \lesssim 10$ mHz,
these \emri{}s will have typical eccentricities $e<0.01$, in agreement
with \M05{}. 

\section{Surviving Binaries}
\label{sec:survive}

We now turn the attention to the population of binaries that survive the encounter
with the \smbh{}. These binaries make up the majority of the encounters we simulated given the 
range of values of $\beta$ we considered. In these scenarios, the energy exchange is simpler: energy is either drained from the orbit of the binary around the \bh{} and
donated to the \CO{} binary orbit, or vice versa (see Eq.~\ref{eq:CM}). In general terms, we expect 
the binary to soften and become bound to the \smbh{} or to tighten and remain unbound.  
Specifically, the surviving binaries will belong to one of the following types: 
binaries bound to the \smbh{} with a Peters lifetime $\tau_{\rm gw}$ longer that their orbital period $P_{\bullet}$ about the hole (i.e. long \bemri{s}); binaries also bound to the \smbh{} but that merge before completing an orbit around the hole, $\tau_{\rm gw}<P_{\bullet}$ (i.e. short \bemri{s}); and binaries that are unbound from the \smbh{.} 
We will analyse the
distributions of orbital parameters (eccentricity and semi-major axis) of the surviving binaries, with
the goal of determining the net change to merger lifetime and whether
this leads to detectable changes in the \CBC{} rate observed by detectors
such as LIGO. But before doing so, we will briefly revisit the effect of  $\theta_0$ and $\Omega_0$.

\begin{figure}
\centering
\includegraphics[width = \columnwidth]{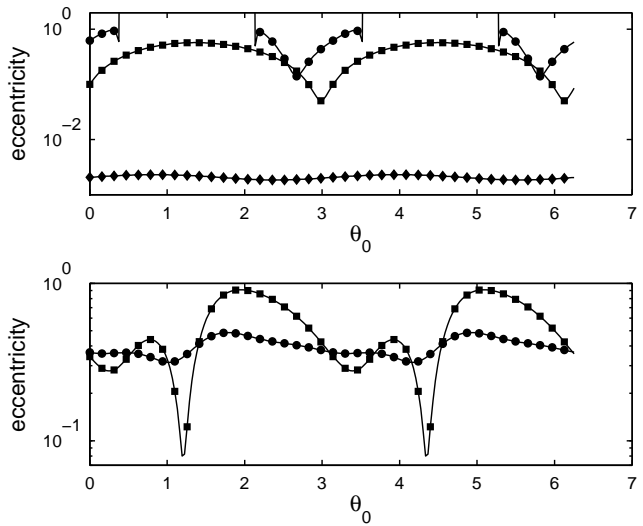}
\caption{Resulting eccentricity of surviving binaries as a function of $\theta_0$. Upper panel: the dots show cases with parameters $(\beta^{-1}, \iota, \Omega_0)=(0.77,123^{\circ},46^{\circ})$, squares with $(1.9,101^{\circ},312^{\circ})$, and  diamonds with $(4.6,49^{\circ},152^{\circ})$. Points where the curve jumps above $1$ indicate disrupted binaries. Lower panel: dots with parameters $(\beta^{-1}, \iota, \Omega_0)=(0.57,177^{\circ},255^{\circ})$, and squares with $(1.7,79^{\circ},308^{\circ})$.}
\label{fig:resultseccTh0_1}
\end{figure}

In Section~\ref{sec:tidal}, we estimated probability distributions of the parameters $\lbrace \beta , \iota, \theta_0, \Omega_0 \rbrace$ for each of the encounter types. We observed that the distributions for the phase of the binary $\theta_0$ and the longitude of the ascending node $\Omega_0$ are to a good approximation flat. Thus, the parameters $\beta$ and $\iota$ have a more dominant  role on the outcome.   This does not mean that choices of $\theta_0$ and $\Omega_0$ do not influence the end state.  Figure \ref{fig:resultseccTh0_1} displays the
eccentricity of the \CO{} binary after the encounter with the \smbh{} as a function of $\theta_0$.
In the upper panel, we show with dots cases with parameters $(\beta^{-1}, \iota, \Omega_0)=(0.77,123^{\circ},46^{\circ})$,  squares with $(1.9,101^{\circ},312^{\circ})$, and  diamonds with $(4.6,49^{\circ},152^{\circ})$. Points where the curve jumps above $1$ indicate disrupted binaries. Notice as expected that as the penetration factor increase, so the eccentricity gained by the binary as well as its dependence of the initial binary phase. In the lower panel in Figure \ref{fig:resultseccTh0_1} we show with circles  parameters $(\beta^{-1}, \iota, \Omega_0)=(0.57,177^{\circ},255^{\circ})$  and with  squares have $(1.7,79^{\circ},308^{\circ})$. Notice that binaries making a closer pass (circles) generally suffers a reduced perturbation to eccentricity due to the nearly retrograde orbit.

\subsection{Post-Encounter Binary Eccentricity}
\label{sec:ecc}
\begin{figure}
\centering
\includegraphics[width = \columnwidth]{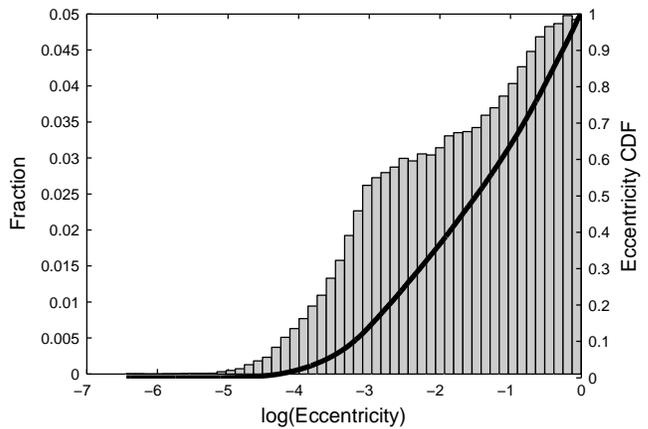}
\caption{Fraction of  surviving binaries as a function of post-encounter eccentricity as well as 
with a solid line the cumulative distribution function (CDF) of the eccentricity values.}
\label{fig:resultsEccHist}
\end{figure}

In Figure~\ref{fig:resultsEccHist}, we show the fraction of  surviving binaries as a function of post-encounter eccentricity as well as 
the cumulative distribution
function (CDF) of the eccentricity values.  Note that the majority of
the surviving binaries are relatively unperturbed in eccentricity,
with the $65\%$ quantile lying at approximately $e\approx
0.109$ and the $90\%$ quantile at $e\approx 0.553$.

\begin{figure}
\centering
\includegraphics[width = \columnwidth]{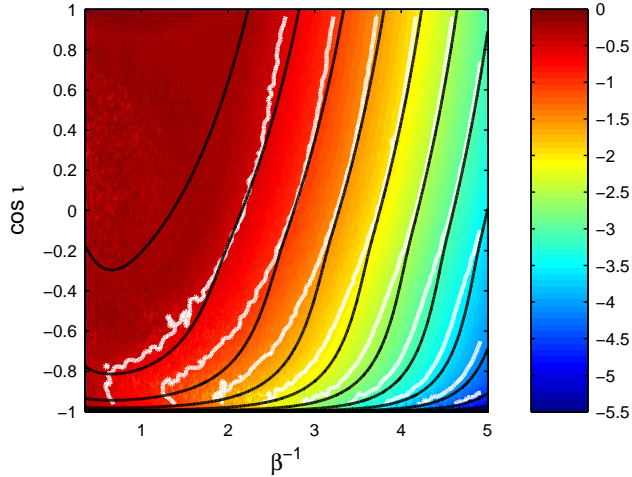}
\caption{Resulting eccentricity $\log e$ for surviving binaries plotted as an intensity map vs. $\beta^{-1}$ and $\iota$. Contours of constant $\log  e =  -5.5,-5.0,-4.5,...,0$ are shown from right to left  in white and for the analytic estimate from Eq.~\ref{eqn:H&Rbeta} in black.}
\label{fig:resultsEccHeatAnal}
\end{figure}

Figure \ref{fig:resultsEccHeatAnal} show how the post-encounter eccentricity depends on $\beta^{-1}$ and $\iota$.
The surviving binary acquires eccentricity for larger penetration factors (smaller $\beta^{-1}$) and prograde orbits ($\cos\iota > 0$). Contours of constant  $\log e = -5.5,-5.0,-4.5,...,0$ from right to left are shown in white. Black lines are the corresponding estimates obtained from Eq.~\ref{eqn:H&Rbeta} averaged over $\phi$. Notice that the contour for $\log e = 0$ is not present for the simulation data. All simulation values have been averaged over $\theta_0$. As expected, we see better agreement between the prediction from Eq.~\ref{eqn:H&Rbeta} and simulation results for large $\beta^{-1}$ and small $\iota$ since that is the regime where Eq.~\ref{eqn:H&Rbeta} was derived. The small overestimation of the prediction at small inclination is consistent with what was found in \citet{1996MNRAS.282.1064H} for an equal mass binary.

\begin{figure*}\centering
\includegraphics[width = \textwidth]{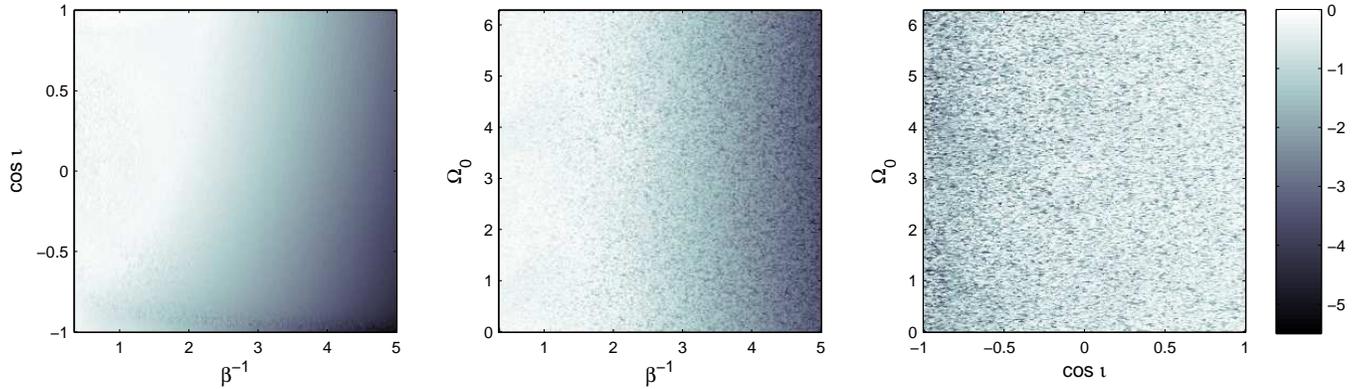}
\caption{Gray scale maps of post-encounter $\log e$ as a function of combinations of input parameters. All values have been averaged over $\theta_0$.}
\label{fig:resultsEccHM}
\end{figure*}

Finally, Figure \ref{fig:resultsEccHM} shows gray scale maps of the post-encounter $\log e$ against  pairs of parameters  $\beta^{-1}, \iota$, and $\Omega_0$, averaged over
$\theta_0$.  Noticeable are the different effects of the input parameters.  As $\beta^{-1}$ increases, the final eccentricity
decreases toward zero, however even binaries with
 $\beta^{-1}=3$ can receive a non-negligible amount of
eccentricity, $e \sim 0.2$, after the encounter.  Inclination plays a
smaller but important role in determining the allowable range of final
eccentricities, with eccentricity decreasing as the binary orbit
approaches retrograde with respect to the \smbh{} orbit.  As stated before, the longitude
of the ascending node $\Omega_0$ plays no predictable role in the
perturbed eccentricity.

\subsection{Post-Encounter Binary Semi-major Axis}

\begin{figure}
\centering
\includegraphics[width = \columnwidth]{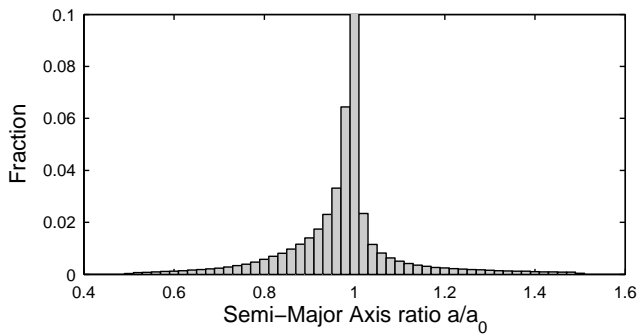}
\caption[Histogram of the binary semi-major axis distribution after \smbh{} encounter.]{Histogram of the binary semi-major axis distribution after \smbh{} encounter. The horizontal axis has been stretched so that the few outliers did not dominate the plot scale, and the vertical axis has been truncated to show detail with the central bin having a true value of $\sim 0.69$. It is clear that the majority of encounter result in very small change in semi-major axis.}
\label{fig:resultsSmaHist}
\end{figure}

Semi-major axis of the surviving binaries is shown as a histogram in
Figure \ref{fig:resultsSmaHist}.  The semi-major axis has been normalized to its initial value. 
It is clear that the semi-major axis is generally
not perturbed to a strong degree, with the overall mean being $\langle a/a_0\rangle =
1.03$.  If the sample size is restricted to binaries with final
semi-major axes in the range $0.5 < a/a_0 < 1.5$, which accounts for
$>97\%$ of the data and removes the outliers, then the mean is
$\langle a/a_0 \rangle = 0.988$ and standard deviation of $0.084$.

\begin{figure*}
\centering
\includegraphics[width = \textwidth]{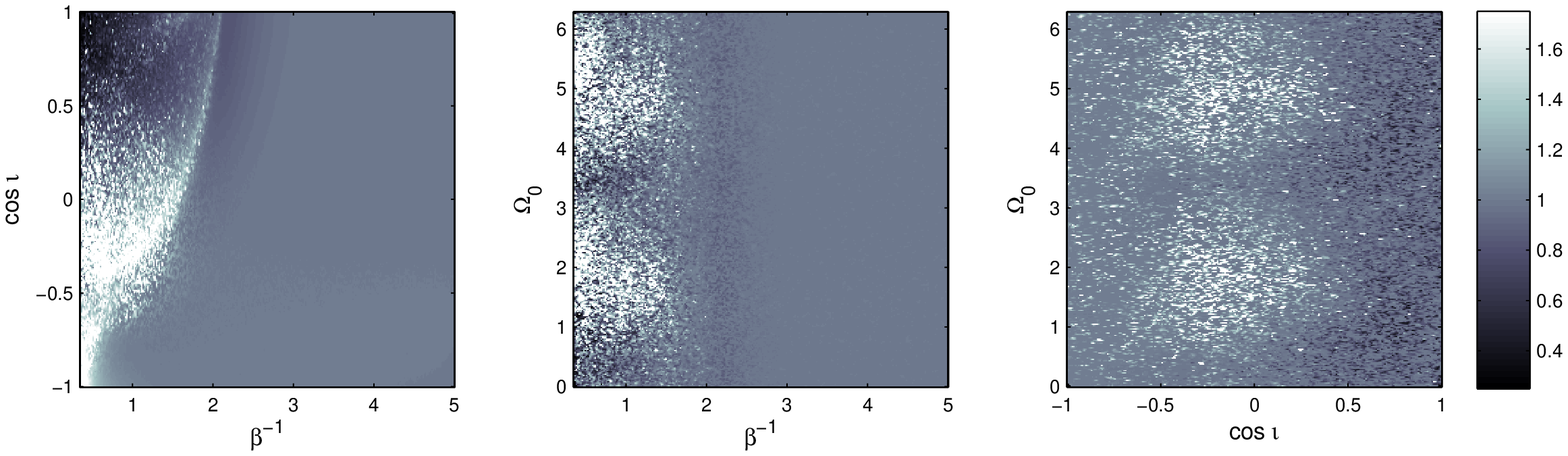}
\caption{Resulting semi-major axis $(a/a_0)$ for surviving binaries plotted as intensity maps vs. pairs of input parameters. All values have been averaged over $\theta_0$.}
\label{fig:resultsSmaGammaInc}
\end{figure*}

Figure \ref{fig:resultsSmaGammaInc} is similar to Figure \ref{fig:resultsEccHM} but for $a/a_0$.  It can be seen
here that for $\beta^{-1} < 2$ the binary can suffer a large change to
the semi-major axis, while above this value the change is small.
Inclination has the general effect of tightening the binary for $\iota
< \pi/2$, and loosening it for $\iota > \pi/2$.  There is some
variation in the extent of semi-major axis perturbation as a function
of $\Omega_0$, particularly when the inclination is around $\iota
\approx \pi/2$ or at low values of $\beta^{-1}$, with minimum change
at $\Omega_0 \approx 0$ or $\pi$.

Since the binding energy of the binary is related to the semi-major as
$E_{b} \propto -a^{-1}$, one can from Figure~\ref{fig:resultsSmaGammaInc}  identify regions of parameter space where the binding energy of the binary increases or decreases.  Furthermore, given that 
$E  = E_{cm} + E_{b} = E_{b,0}  <  0$, an increase in binary energy
$E_{b}$ must be compensated by a decrease in the \bh{} orbit
energy $E_{cm}$, resulting in capture of the binary.  Therefore the
points in Figure \ref{fig:resultsSmaGammaInc} where $a/a_{0} > 1$
correspond to binaries which become bound, while $a/a_{0} < 1$
indicates binaries that remain unbound.

\section{Peters Lifetime and CBC Rates}
\label{sec:Peters}

\begin{figure*}
\centering
\includegraphics[width = \textwidth]{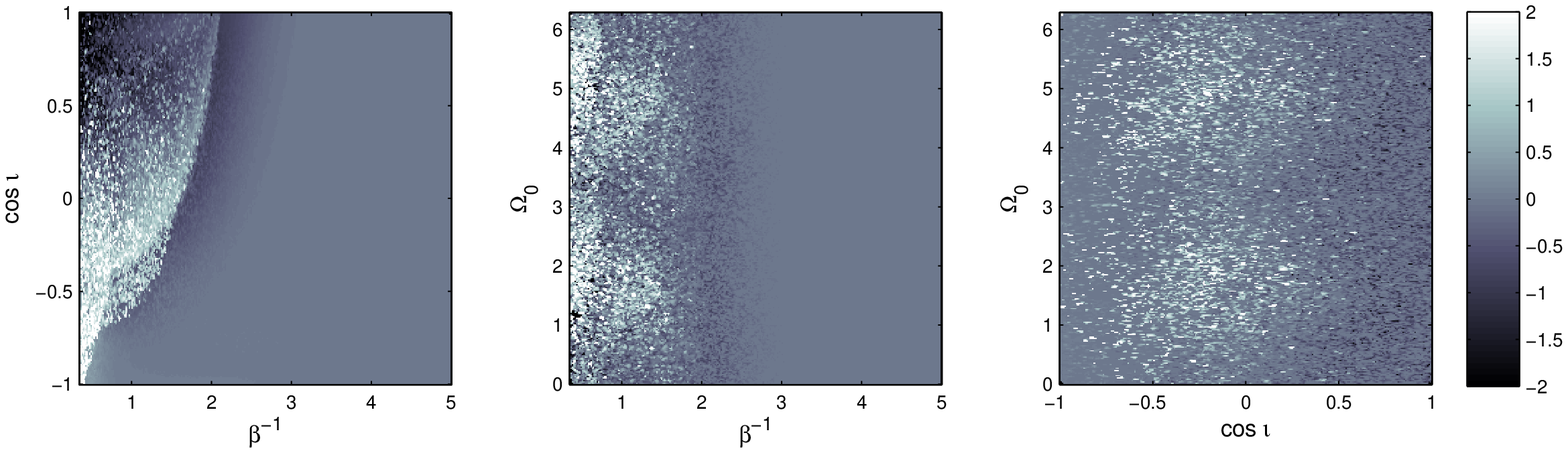}
\caption{Gray scale maps of the Peters lifetime $\log(\tau_{\rm gw}/\tau_{\rm gw,0})$ for surviving binaries vs. pairs of input parameters. All values have been averaged over $\theta_0$.}
\label{fig:resultsLifGammaInc}
\end{figure*}

As mentioned before, the Peters lifetime is an estimate of the time that a binary system with eccentricity $e$ and semi-major axis $a$ takes to merge as it looses energy due to emission of \gw{s}~\citep{1964PhRv..136.1224P}.  
As presented in Equation~\ref{eq:pedro}, the \CO{} binary system in our study, with initial semi-major-axis $a_0 = 10\,R_\odot$ and vanishing eccentricity (i.e. circular orbit),  has a Peters lifetime of $\tau_{{\rm gw},0} \approx 10^8$ years.  
Figure \ref{fig:resultsLifGammaInc} shows the Peters lifetime $\tau_{\rm gw}$ normalized to $\tau_{\rm gw,0}$  for all of the surviving binaries.  Notice the similarity of the maps with those for the semi-major axis in Figure~\ref{fig:resultsSmaHist}.

Unbound binaries that survive the encounter with the \smbh{,} what we call SU binaries, could in principle have a shorter  ($\tau_{\rm gw} < \tau_{\rm gw,0}$) or longer ($\tau_{\rm gw} > \tau_{\rm gw,0}$) Peters lifetime.
The last situation, however, is not possible given our initial setup.
An unbound binary requires increasing the energy of its center of mass $E_{\rm cm}$. This will come at the expense of decreasing its binding energy
$E_b$ since $E  = E_{cm} + E_{b} = E_{b,0}  <  0$, which in turn requires decreasing its semi-major axis.  A decrease in its semi-major axis translates into a shorter Peters lifetime.  Thus, all of the unbound
binaries in our study will have accelerated merger times due to the \smbh{}

We recall that a {\it short \emri{}} is one for which $\tau_{\rm gw}<P_{\bullet}$. That is, the binary is expected to merge before returning to the hole. On the other hand, a {\it long} \bemri{} is one in which $\tau_{\rm gw} >P_{\bullet}$.  
These \bemri{s} have typically highly elliptical orbits with average eccentricity
$e_{\rm cm} \approx 1-10^{-5}$ and long periods. 
Therefore,
\begin{eqnarray}
P_{\bullet} &=& \frac{2\,\pi\,a_{\rm cm}^{3/2}}{\sqrt{G\,M_\bullet}}= \frac{2\,\pi\,r_p^{3/2}}{\sqrt{G\,M_\bullet}}(1-e_{\rm cm})^{-3/2}\nonumber\\
 &=& \beta^{-3/2}(1-e_{\rm cm})^{-3/2}\frac{2\,\pi\,r_t^{3/2}}{\sqrt{G\,M_\bullet}}\nonumber\\
 &=& \beta^{-3/2}(1-e_{\rm cm})^{-3/2}\frac{2\,\pi\,a_0^{3/2}}{\sqrt{G\,M_b}}\nonumber\\
&=& 7.2\times 10^4\,{\rm yr}\,\beta^{-3/2}\notag\\
&\times&\left(\frac{1-e_{\rm cm}}{10^{-5}}\right)^{-3/2}\,\left(\frac{P_{b,0}}{7.16\times 10^4\,{\rm s}}\right)\label{eq:pbullet}
\end{eqnarray}
The \CO{} binary in a long \bemri{} could potentially survive several orbits
around the \smbh{,} with its orbit being perturbed by each passage. We will investigate the fate of \bemri{} systems
will be the focus of a future study.

Independently of being in an \bemri{} or unbound, \CO{} binaries that after an encounter with a \smbh{} are affected in such a way that $\tau_{\rm gw} < \tau_{\rm gw,0}$ have particular interest. 
They could potentially increase the rates of \CBC{} to be detected by LIGO. 
We construct a simple formula to obtain a first estimate of this effect as
\begin{equation}
\label{eqn:enhance}
\mathcal{E}_{\CBC{}} = [ \Gamma *f_b* N_{G}(D_{h})]*(\mathcal{E}_{T}*f_L),
\end{equation}
where $\mathcal{E}_{\CBC{}}$ is the enhancement to the current predicted \CBC{} rate $R_{\CBC{}}$, i.e., $\hat{R}_{\CBC{}} = R_{\CBC{}} + \mathcal{E}_{\CBC{}}$. The first three terms in Eq. \ref{eqn:enhance} are observational quantities taken from the literature. $\Gamma$ is the estimated encounter rate of stellar mass objects / per \smbh{} per year, $f_b$ is the binary fraction, and $N_{G}(D_{h})$ is the number of Milky Way equivalent galaxies (MWEGs) observable by a \gw{} detector with horizon distance, $D_{h}$. The product of these factors is the rate of binary encounters with a \smbh{} in the observable volume of a \gw{} detector.

The next two factors come from our simulation results. The single binary enhancement $\mathcal{E}_{\tau} = (1 - {\tilde{\tau}/\tau_0})$ is the percent difference between the old merger lifetime, $\tau_{0}$, and the mean new lifetime, $\tilde{\tau}$, and $f_L$ is the fraction of binaries from our simulations that result in guaranteed LIGO sources after the \smbh{} encounter. 
Encounter rates in galaxies with a $10^{6} M_{\odot}$ \smbh{}, estimates for the single \CO{} capture rate range from $\Gamma \sim 5\times 10^{-9} \text{ yr}^{-1} \text{ MWEG}^{-1}$ to as high as $\Gamma \sim 10^{-6} \text{ yr}^{-1} \text{ MWEG}^{-1}$ (\cite{1995ApJ...445L...7H,1997MNRAS.284..318S,2002MNRAS.336.373I,2005ApJ...629..362H,2011PhRvD..84d4024M}); we use the higher side of these estimates, $\Gamma = 10^{-7} \text{ yr}^{-1} \text{ MWEG}^{-1}$, as the pericenter distances in our simulations reach much larger values than the single star capture radius. The binary fraction, $f_b$, near the galactic center is not a well determined value, however, in the absence of better knowledge, we take it to be roughly the same as the binary fraction of field stars, $f_b=0.5$. The number of MWEGs observable by aLIGO with a BH-BH merger horizon distance of $D_{h} = 2187$ Mpc  is given as (\cite{2010CQGra..27q3001A, 0004-637X-556-1-340}) 
\begin{equation}
N_{G} = \dfrac{4}{3} \pi \left ( \dfrac{D_{h}}{\text{Mpc}}\right )^{3}\dfrac{0.0116}{(2.26)^{3}},
\end{equation}
which gives a value of $N_{G} \approx 4.4\times10^{7}$ MWEGs. 

From our simulation results, we find that binaries from the $f_L$ categories have a mean Peters lifetime of $\tilde{\tau} \approx 0.84\tau_{0}$ giving $\mathcal{E}_T \approx 0.16$ and LIGO fraction $f_L \approx 0.68$. 
Combining these factors, we compute an estimated enhancement to the \CBC{} rate of $\mathcal{E}_{\CBC{}} \approx 0.25 \text{ yr}^{-1}$. The predicted rate of expected BH-BH mergers has been estimated to lie between $0.4 \text{ MWEG}^{-1}\text{ Myr}^{-1} < \Gamma_{BH} < 30 \text{ MWEG}^{-1}\text{ Myr}^{-1}$ for realistic to optimistic scenarios \cite{2007PhR...442...75K,2010CQGra..27q3001A}. This corresponds to an estimated merger rate within the aLIGO volume of $\sim 20 \text{ yr}^{-1}< \dot{N}_{BH} < 1300 \text{ yr}^{-1}$, which our estimated rate enhancement $\mathcal{E}_{\CBC{}}$ may change by as much as $\approx 1 \%$. This enhancement will likely be difficult to detect with small observation catalogs and given the uncertainty in the estimated merger rates, but could become noticeable over long observation times.

\section{Summary and Conclusions}
\label{sec:disc}

In this paper, we have presented the results of $\sim 13$ million
individual simulations of parabolic encounters between a \CO{} binary
 and a galactic center  \smbh{} while varying the orientation of the binary
and its distance of closest approach to the \smbh{}.

Tidal disruption of the binary occurs with about $16 \%$ probability
given the full range of parameters covered in this study.  Consistent with previous work in this
area, this sort of disruption can create \HVS{} which can escape from the
\smbh{} with high speed.  We also explored disruption as a formation
mechanism for \emri{}s, which are of interest to space-based \gw{} detection
missions, and found that the \emri{}s formed in this way will generally
have very low eccentricity when they enter the LISA band.  This work
shows that considering the full range of possible orientations gives a
broader range of formation eccentricities than previous estimates have
predicted.

Surviving binaries can either become bound to the \smbh{} after the
encounter and become a \bemri{} or remain unbound from the hole.  
Among them, there is a subclass of surviving binaries for which
$\tau_{\rm gw}<\tau_{\rm gw,0}$  those with a small enough \bh{} orbital
period such that they will not merge before completing one orbit are
the ``short period" B\emri{}s, and understanding their full evolution
requires a more careful (i.e. Post-Newtonian) approach to the
integration in order to account for eccentricity and semi-major axis
change due to \gw{} loss during the long orbit.  Both the unbound
binaries and the long period B\emri{}s which will merge before one \smbh{}
orbit are factored into the calculation for the \CBC{} rate enhancement,
which is potentially important to ground based \gw{} detectors such as
LIGO. We find that for the aLIGO volume, the enhancement factor for
BH-BH mergers is $\mathcal{E}_{\CBC{}} \approx 0.25 \text{ yr}^{-1}$ or as
much as $1\%$ of the predicted rates, though this enhancement may be
difficult to statistically detect in accumulated event catalogs.

\acknowledgements
PL supported by NSF grants 1205864, 1212433, 1333360. EA and SLL supported by NSF grant
PHY-0970152, and from NASA award NNX13AM10G.

\bibliography{Busting_Up_Binaries}

\end{document}